\documentclass[]{interact}

\usepackage{epstopdf}
\usepackage{subfigure}

\usepackage{natbib}
\bibpunct[, ]{(}{)}{;}{a}{}{,}

\usepackage{amsthm,amsfonts,amssymb,amsmath}
\usepackage{xspace}
\usepackage{bm}
\usepackage{graphicx}

\newcommand{\R}{\mathbb{R}}

\newcommand{\bx}{\mathbf{x}}

\newcommand{\bydef}{:=}

\theoremstyle{plain}

\theoremstyle{definition}

\theoremstyle{remark}

\begin{document}


\title{Evolutionary Game Dynamics Applied to Strategic Adoption of Immersive Technologies in Cultural Heritage and Tourism}

\author{
\name{Gioacchino Fazio\textsuperscript{a}, Stefano Fricano\textsuperscript{a}\textsuperscript{*}\thanks{\textsuperscript{*}CONTACT Stefano Fricano. Email: stefano.fricano@unipa.it} and Claudio Pirrone\textsuperscript{a}}
\affil{\textsuperscript{a}Department of Economics, Business and Statistics, University of Palermo,  Viale delle Scienze Ed.13, Palermo, IT-90128}
}

\maketitle

\begin{abstract}
Immersive technologies as Metaverse, AR, and VR are at a crossroads, with many actors pondering their adoption and potential sectors interested in integration. The cultural and tourism industries are particularly impacted, facing significant pressure to make decisions that could shape their future landscapes. Stakeholders' perceptions play a crucial role in this process, influencing the speed and extent of technology adoption. As immersive technologies promise to revolutionize experiences, stakeholders in these fields are weighing the benefits and challenges of embracing such innovations. The current choices will likely determine the trajectory of cultural preservation and tourism enhancement, potentially transforming how we engage with history, art, and travel. Starting from a decomposition of stakeholders' perceptions into principal components using Q-methodology, this article employs an evolutionary game model to attempt to map possible scenarios and highlight potential decision-making trajectories. The proposed approach highlights how evolutionary dynamics lead to identifying a dominant long-term strategy that emerges from the complex system of coexistence among various stakeholders.
\end{abstract}


\section{Introduction}

Immersive technologies are at a pivotal juncture, as numerous sectors contemplate their adoption and potential applications (\cite{Monaco2023}). Among these sectors, cultural and tourism industries are particularly impacted, facing significant pressure to decide whether and how to integrate these immersive technologies (\cite{Koo2022}, \cite{Yang2023}, \cite{Volchek2023},\cite{Shin2024}). The rapid development of Immersive technologies presents both opportunities and challenges for these sectors. Museums, galleries, and heritage sites can harness these immersive technologies to create engaging, interactive experiences that attract new audiences and deepen visitor engagement (\cite{wedel2020}). 

Immersive technologies are a broad spectrum of advanced digital tools and systems that create or enhance the sense of physical presence in a virtual or simulated environment. They represent the intersection of the physical and digital worlds, where users can interact with and experience environments that feel as though they are a part of them, rather than merely observing them. These technologies leverage a combination of hardware and software to generate environments that can be entirely artificial or an augmentation of the real world.

One of the most recognized forms of immersive technology is Virtual Reality (VR). VR is a fully immersive experience where the user is placed in a completely digital environment. This is achieved through devices such as headsets, which cover the user’s field of vision and often include headphones to block out real-world sensory input. In VR, users are transported to a different world, one where they can look around in 360 degrees, move within the space, and interact with objects as if they were physically present. The key characteristic of VR is that it creates an entirely new reality for the user, detaching them from their physical surroundings.

In contrast, Augmented Reality (AR) enhances the real world by overlaying digital information onto the physical environment. Unlike VR, AR does not create a new world but enriches the existing one with additional layers of information. This can be achieved through devices like smartphones, tablets, or AR glasses that project digital elements such as images, sounds, or data over the user's view of the real world. For example, AR might allow a person to see directions on the road in front of them, or visualize how a piece of furniture would look in their living room before purchasing it. The distinction between AR and VR is crucial: AR augments reality, while VR replaces it.

Mixed Reality (MR) is another form of immersive technology that blends aspects of both VR and AR. In MR, digital objects are not just overlaid on the real world but are anchored to it and interact with it in real time. This means that a virtual object in MR can be occluded by real-world objects, or respond to changes in lighting and environment. MR creates a seamless integration between the physical and digital worlds, where both can coexist and interact with each other in meaningful ways. A key example of MR is Microsoft’s HoloLens, a device that allows users to see and interact with holograms within their real environment.

Lastly, Extended Reality (XR) is a term that encompasses all forms of immersive technologies, including VR, AR, and MR. XR serves as a broad umbrella under which these technologies fall, representing the entire spectrum of digitally mediated experiences. XR is often used to describe the evolving landscape of immersive technologies as they continue to develop and converge, pushing the boundaries of what is possible in digital experiences.

Despite their similarities, these technologies offer distinct experiences. VR fully immerses users in a new world, AR enhances the real world with digital overlays, MR creates a blend of real and virtual interactions, and XR encompasses all these forms, representing the future of immersive digital experiences. Each has its own set of use cases, advantages, and challenges, making them suitable for different applications across industries such as gaming, education, healthcare, and more.

For the tourism industry in particular, AR and VR can offer virtual tours, enhancing marketing efforts and providing unique experiences to potential travellers (\cite{Yang2023}). Immersive technologies promise to revolutionize experiences by offering unprecedented levels of engagement and interaction (\cite{Williams1995}, \cite{Lee2011},  \cite{Buhalis2023}). In the cultural sector, museums and galleries could use AR and VR to create more interactive and informative exhibits, enhancing visitors' understanding and enjoyment of art and history (\cite{Gursoy2022}). 

Similarly, the tourism industry could leverage these technologies to offer virtual tours of landmarks and destinations, allowing potential travellers to explore places before visiting in person (\cite{McKercher2020}). 

However, the integration of these technologies is not without challenges. High costs, technical complexities, and concerns about accessibility and inclusivity must be addressed. Stakeholders need to carefully consider these factors and weigh them against the potential benefits. Stakeholders' perceptions play a crucial role in this decision-making process, shaping the pace and extent of adoption (\cite{Clarkson1995}, \cite{Freeman2008}). 

So, the cultural and tourism industries are at a critical juncture, experiencing significant pressure to make decisions that could shape their future landscapes (\cite{baran2022}). 
The current decisions will likely determine the future trajectory of cultural preservation and tourism enhancement. 

Stakeholders' perceptions are pivotal in determining the speed and extent of technology adoption, particularly in the integration of the metaverse, AR, and VR within various sectors (\cite{Fazio2024}). In the cultural and tourism industries, these perceptions can significantly influence whether these technologies are embraced or resisted. Stakeholders include a diverse group—ranging from industry leaders and policymakers to consumers and technology developers—each bringing unique perspectives and concerns (\cite{Freeman2008}).
For instance, cultural institutions like museums may perceive AR and VR as valuable tools for enhancing visitor engagement and education, leading to a more enthusiastic adoption. Conversely, concerns about high implementation costs, technical challenges, and potential disruptions to traditional experiences might slow down their integration (\cite{jia2023}). Similarly, in the tourism sector, stakeholders might view immersive technologies as revolutionary for marketing and virtual exploration, driving quicker adoption. Yet, scepticism about their long-term value and the readiness of the market could temper this enthusiasm (\cite{Gursoy2022}, \cite{Karaca2023}).

Effective communication and collaboration among stakeholders are essential to address these diverse perceptions. Providing clear evidence of the benefits, addressing technical and financial concerns, and ensuring inclusivity and accessibility can help align stakeholders' views. Moreover, pilot projects and case studies demonstrating successful implementations can build confidence and support. Ultimately, stakeholders' perceptions shape the trajectory of technology adoption, impacting how quickly and extensively AR, VR, and the metaverse are integrated (\cite{Buhalis2023}).

Stakeholders such as government bodies, private enterprises, cultural institutions, and consumers bring unique perspectives and incentives, which contributes to the complexity of analysing their interactions. Each stakeholder type influences and is influenced by others, creating a network of interdependencies. Understanding all possible interactions is challenging due to the varied objectives, risk tolerances, and strategic behaviours involved. Each stakeholder group typically has its own set of objectives, resources, and constraints, which can vary widely and change over time. Additionally, the relationships between stakeholders are often non-linear and can involve complex feedback loops, where the actions of one group affect others in unpredictable ways. This complexity poses significant challenges for researchers attempting to model these interactions in a meaningful way.

In the scientific literature, the problem of stakeholder interdependence has often been under-explored, primarily due to the inherent complexity of the issue and the challenges associated with developing suitable models for its study. These interconnections can significantly influence decision-making processes and outcomes, making it a critical area of study for understanding how organizations and systems operate within larger networks.

The lack of robust models to study stakeholder interdependence has led to a scarcity of research in this area, despite its importance. Addressing this gap requires the development of more sophisticated, multi-dimensional models that can better capture the nuances of stakeholder interactions.  This complexity necessitates, in our opinion, sophisticated modelling approaches, such as game theory, to predict outcomes and devise effective strategies. By modelling these interactions as games, researchers can identify potential strategies stakeholders might adopt and the possible outcomes of their decisions. 

In fact, game theory can help to map out stakeholders' preferences and predict which strategies will dominate over time (\cite{Taylor1978}). Game theory provides a mathematical framework to analyse strategic interactions where the outcome for each participant depends on the choices of all involved (\cite{Smith1974}). The replicator dynamics of game theory considers how strategies evolve over time based on their success relative to others (\cite{Ferriere2011}). In the context of innovation adoption, the replicator dynamic functions by comparing the success of individuals who adopt a new innovation to those who do not. The idea is that if an innovation provides a higher payoff—be it economic gain, increased efficiency, improved social standing or, generally speaking, higher utility—then the individuals who adopt it will have a relative advantage over those who stick to older methods. Over time, as these individuals succeed, the innovation is expected to spread through the population, much like a successful strategy spreads in an evolutionary game.

Scientific studies often model this process by considering a population where individuals have the choice between adopting a new innovation or sticking with a traditional approach. The payoffs associated with each choice depend on several factors, such as the cost of adoption, the benefits derived from the innovation, and the proportion of the population that has already adopted the innovation. The replicator dynamic then describes how the proportion of adopters changes over time. If the innovation proves to be more successful, the proportion of adopters increases, potentially leading to the innovation becoming widespread or even universally adopted within the population.

One of the key insights from applying replicator dynamics to innovation adoption is the understanding of equilibrium points—situations where the proportion of the population adopting the innovation stabilizes. These equilibrium points can be stable or unstable, depending on the nature of the innovation and the environment. A stable equilibrium suggests that once the innovation reaches a certain level of adoption, it will continue to be used by the population, while an unstable equilibrium might indicate that slight changes in the population's behaviour could lead to either widespread adoption or complete rejection of the innovation. Fundamental, in this case, is the perception of the "usefulness" that the actors of the population (stakeholders) have of the innovation and of the "hierarchy" of the various choices.

Decomposing stakeholders' perceptions into common factors and treating the complex system as a statistical ensemble can be invaluable for understanding dynamics. Recent advancements in Q-methodology allow for the identification of key factors shaping stakeholder perceptions (\cite{Fazio2024}). By systematically analysing subjective viewpoints, researchers can distil complex opinions into essential components. This method enables a deeper understanding of stakeholders' perspectives, uncovering underlying patterns and priorities. Through Q-methodology (\cite{stephenson1935}), stakeholders' diverse viewpoints can be organized and categorized, providing valuable insights for decision-making processes (\cite{Gao2020}). This approach enhances our ability to navigate the complexities of stakeholder dynamics.

In this paper, we propose to simultaneously analyse the interactions between a set of strategies and possible combinations of Q-key factors. Through a progressive selection aimed at maximizing perceived utility for stakeholders, we intend to employ evolutionary models to identify patterns and trends.
This method enhances our ability to navigate the complexities of technology adoption in diverse sectors and highlights how certain strategies, such as early adoption of immersive technologies, can become dominant through a progressive selection among different strategies. 
These models help to clarify the likely paths of technology adoption and the conditions under which stakeholders might cooperate or compete. By understanding these dynamics, policymakers and industry leaders can devise better strategies to foster collaboration, mitigate conflicts, and promote the effective integration of new technologies, ultimately benefiting the entire ecosystem (\cite{Pappu2024}).

\section{Materials and methods}

\subsection{Q-meth and stakeholders decomposition}
\label{qmeth}
In \cite{Fazio2024}, they examined the potential impact of metaverse technologies on the tourism sector. The study focused on understanding how different tourism stakeholders, such as managers of attractions like archaeological parks, museums, and nature reserves, perceive and integrate metaverse tools into their strategies.

Key findings indicate that stakeholders have varied preferences and levels of awareness regarding the use of metaverse technologies. Some are fully aware and view these tools as beneficial, while others lack sufficient knowledge and remain hesitant. This variance suggests that the tourism sector is still in a transitional phase concerning the adoption of metaverse technologies. 

The results showed that the diffusion of metaverse in tourism is influenced by several factors, including planning, management, economic and sociocultural factors, as well as technological acceptance. They concluded that while there is enthusiasm about the potential benefits, there are also significant challenges, particularly in achieving widespread adoption and understanding among all stakeholders involved.

The study applied a recent development of Q-methodology which allows substituting classical q-statements formulation with multi-attribute and multi-level formulations (\cite{Gao2020}). Data are collected by involving stakeholders from Sicilian territories. In this paper, we use the results obtained in \cite{Fazio2024}, based on the highlighted findings, to understand what possible future scenarios might unfold.

\subsubsection{Q-meth results summary}
\label{qresult}
In their paper, \cite{Fazio2024} have decided to focus their analysis on the following four main characteristics that they recognize to concur with the definition of the best tourism marketing strategy:
\begin{itemize}
	\item Main target;
	\item Main content;
	\item Principal tool (or media channel);
	\item Preferred financial scheme (resources).
\end{itemize}

Referring to the first item, they decided to differentiate it into two options: domestic and international tourists. Regarding the second one, they considered two possibilities: content purely for recreational entertainment vs. content aimed at cultural growth. They assumed, as a media channel to develop attractiveness, three options: traditional advertising tools as standard media (TV or print advertising), consolidated digital tools (Internet/social-media advertising), or advanced digital tools such as Metaverse (AR, VR). Lastly, they referred to economic resources to identify the main actors implementing the strategy. We provided three options: public resources, private resources, or a mixed public/private resource.

So, participants were provided with different levels of the four variables:\textbf{ Target, Content, Tool, and Resource}.

\begin{table}[htbp!]
	\begin{center}
		\begin{tabular}{|c|c|c|c|}
			\hline
			Target & Content & Tool & Resource \\
			\hline
			Domestic (D) & Recreational (R) & Traditional (T) & Public (Pu) \\
			International (I) & Cultural (C) & Standard digital (S) & Private (Pr) \\
			& & Advanced digital (A) & Mixed (PP) \\
			\hline
		\end{tabular}\\
	
		\caption{Available options for each of the dimensions that contribute to forming the possible strategies.} \label{codifica}
	\end{center}
\end{table}
A typical statement had the following form:
\
\\
\begin{center}
	\textit{
		For me, the best strategy to increase the tourist attraction of the place where I live is the one that aims at an \textbf{international tourist target}, based on a proposal of purely \textbf{recreational content} that uses \textbf{traditional tools (such as TV or print advertising)} and funded by \textbf{public economic resources}}.
	\\
\end{center}
\
\\
Combining the different levels of the four variables, they obtained thirty-six statements (they identified the above statement by the code: \textbf{\textit{I.R.T.Pu}} and so on for the other statements).

\begin{figure}[htbp!]%
	\centering
	\includegraphics[width=0.9\textwidth]{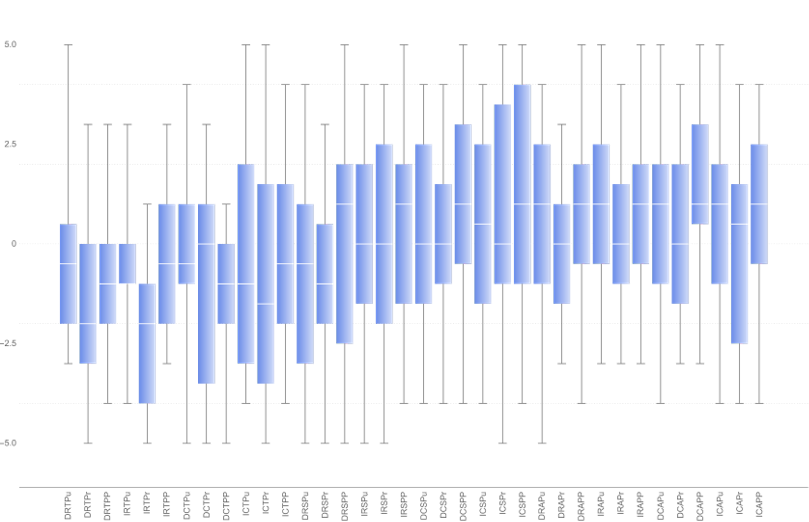}
	\caption{Box-plot of the distributions of the values obtained for each statement (\cite{Fazio2024}).}\label{figura2}
\end{figure}

Following the directions of the Q-sort methodology process (\cite{Watts2012}), Fazio et al. focused on 20 stakeholders of different nature and with different roles. 
As a result of the Q-sort survey carried out, for each of the 36 combinations/statements about the best destination marketing strategy, they obtained 20 evaluations (one from each of the participants) ranging from -5 (in case they consider it the one with which they most disagree) to 5 (in the case consider it the one with which they are most in agreement). In a certain sense, we can imagine that agreeing or disagreeing with individual strategies could represent their perception about the usefulness of each of the strategies.
In Figure \ref{figura2}, they have reported the box-plot of the values obtained for each statement they have recorded. 

They have developed principal component analysis (PCA) with Varimax rotation to analyse the Q-sorting data (\cite{stephenson1935}). The number of Q-factors depends on the data; the usual criteria by which the number of components is selected include the total amount of variability explained or eigenvalues higher than a certain threshold. The chosen procedure of finding statistically significant factors led to considering only the principal components for which the eigenvalue was more significant than 1 (\cite{Lee2017}). This led to the identification of \textbf{5} principal components on which a varimax rotation procedure was performed; these collect almost 80\% of the total variance.

Following the criterion proposed by Zabala (\cite{Zabala2014}), they can "flag" each of the participants with respect to the 5 factors based on loadings that are shows in Table \ref{loadings}.

\begin{table}[htbp!]
	\begin{center}
		\begin{tabular}{|c|c|c|c|c|c|}
			\hline
			Stakeholder & Q-Factor-1 & Q-Factor-2 & Q-Factor-3 & Q-Factor-4 & Q-Factor-5 \\   
			\hline
			STK1 & \textbf{0,84} & 0,17 & 0,07 & -0,12 & 0,27 \\   
			\hline
			STK2 & 0,67 & -0,18 & 0,01 & 0,15 & 0,64 \\   
			\hline
			STK3 & 0,07 & 0,53 & 0,31 & \textbf{0,7} & -0,08 \\   
			\hline
			STK4 & \textbf{0,79} & -0,39 & -0,2 & 0,06 & -0,14 \\   
			\hline
			STK5 & \textbf{0,67} & -0,16 & 0,13 & 0,08 & 0,25 \\   
			\hline
			STK6 & \textbf{0,88} & 0 & 0,06 & -0,13 & -0,05 \\   
			\hline
			STK7 & \textbf{0,85} & 0,04 & 0,04 & -0,04 & -0,12 \\   
			\hline
			STK8 & \textbf{0,68} & -0,15 & 0,1 & 0 & 0,65 \\   
			\hline
			STK9 & -0,29 & 0,13 & -0,28 & 0,29 & \textbf{-0,66} \\   
			\hline
			STK10 & 0,02 & 0,1 & \textbf{0,95} & 0,05 & 0,15 \\   
			\hline
			STK11 & \textbf{0,61} & -0,48 & -0,29 & -0,06 & -0,13 \\   
			\hline
			STK12 & 0,02 & 0,1 & \textbf{0,95} & 0,05 & 0,15 \\   
			\hline
			STK13 & 0,22 & -0,1 & 0,12 & \textbf{0,73} & 0,26 \\   
			\hline
			STK14 & -0,09 & 0,38 & 0,19 & \textbf{0,86} & -0,14 \\   
			\hline
			STK15 & -0,03 & 0,13 & \textbf{-0,6} & -0,18 & 0,46 \\   
			\hline
			STK16 & -0,05 & \textbf{0,81} & 0,18 & 0,05 & -0,08 \\   
			\hline
			STK17 & 0,36 & 0,08 & 0,19 & \textbf{-0,76 }& 0,04 \\   
			\hline
			STK18 & -0,09 & \textbf{-0,69} & 0,12 & -0,18 & 0,06 \\   
			\hline
			STK19 & 0,24 & 0,01 & 0 & -0,18 & \textbf{-0,87} \\   
			\hline
			STK20 & -0,31 & \textbf{0,81} & -0,01 & -0,05 & -0,03 \\   
			\hline
			
		\end{tabular}
	\end{center}
	\caption{Loading values of each participant; highlighted the values used to flag stakeholders on Q-factors (\cite{Fazio2024}). We can assign 7 stakeholder to Q-Factor1 (35\%), 3 to Q-Factorr-2 (15\%), 3 to Q-Factor-3 (15\%), 4 to Q-Factor-4 (20\%) and 2 to Q-Factor-5 (10\%). One stakeholder has not been assigned to any Q-Factor.}\label{loadings}
\end{table}

The Zabala's R-package (\cite{Zabala2014}) provides also a simulation of the answers from the various points of view of the Q-factors by assigning a factored score to each of the statements following the distribution of the values (from -5 to +5); these are listed in Table \ref{conse}. 

\begin{table}[htbp!]
	\small 
	\begin{center}
		\begin{tabular}{|c|c|c|c|c|c|}
			\hline
			& & & & & \\
			Strategies & $zscore^{Q_1}$  & $zscore^{Q_2}$ & $zscore^{Q_3}$ & $zscore^{Q_4}$ & $zscore^{Q_5}$ \\   
			& & & & & \\
			\hline
			D.R.T.Pu. & -3 & -1 & -2 & 2 & 1   \\   
			\hline
			D.R.T.Pr. & -2 & -2 & -2 & -2 & -3  \\   
			\hline
			D.R.T.PP. & -3 & 1 & -2 & 1 & 0  \\   
			\hline
			I.R.T.Pu. & -1 & 0 & -1 & 3 & 1 \\   
			\hline
			I.R.T.Pr. & -4 & -2 & 0 & -2 & -1  \\   
			\hline
			I.R.T.PP. & -2 & 2 & 0 & 0 & -1   \\   
			\hline
			D.C.T.Pu. & -1 & 0 & 4 & -1 & 0 \\   
			\hline
			D.C.T.Pr. & -5 & -3 & 3 & 0 & -2  \\   
			\hline
			D.C.T.PP. & -2 & 3 & -1 & -1 & -1  \\   
			\hline
			I.C.T.Pu. & -3 & 0 & 5 & 3 & 0 \\   
			\hline
			I.C.T.Pr. & -4 & -4 & 3 & -3 & 0 \\   
			\hline
			I.C.T.PP. & -2 & 1 & 2 & 2 & -2  \\   
			\hline
			D.R.S.Pu. & 0 & 2 & -5 & -3 & 2  \\   
			\hline
			D.R.S.Pr. & -1 & -3 & -3 & 0 & 1  \\   
			\hline
			D.R.S.PP. & 2 & 4 & -4 & 1 & 3  \\   
			\hline
			I.R.S.Pu. & -1 & 1 & -3 & -5 & 4  \\   
			\hline
			I.R.S.Pr. & 2 & -2 & -3 & 1 & 3  \\   
			\hline
			I.R.S.PP. & 2 & 3 & -4 & 0 & 2 \\   
			\hline
			D.C.S.Pu. & 3 & 2 & 0 & -4 & -1 \\   
			\hline
			D.C.S.Pr. & 3 & -1 & 1 & 0 & 2  \\   
			\hline
			D.C.S.PP. & 3 & 5 & 0 & 0 & 2  \\   
			\hline
			I.C.S.Pu. & 1 & 1 & 1 & -2 & 3  \\   
			\hline
			I.C.S.Pr. & 4 & -1 & 0 & 2 & 5 \\   
			\hline
			I.C.S.PP. & 1 & 3 & 0 & -1 & 4  \\   
			\hline
			D.R.A.Pu. & 0 & -1 & 1 & 4 & -3  \\   
			\hline
			D.R.A.Pr. & 0 & -2 & -2 & -1 & 0   \\   
			\hline
			D.R.A.PP. & 5 & 4 & -1 & 1 & 1  \\   
			\hline
			I.R.A.Pu. & -1 & -1 & 2 & 5 & -2  \\   
			\hline
			I.R.A.Pr. & 0 & -4 & -1 & -3 & -3   \\   
			\hline
			I.R.A.PP. & 2 & 2 & -1 & 1 & -5  \\   
			\hline
			D.C.A.Pu. & 4 & 0 & 2 & 2 & -4  \\   
			\hline
			D.C.A.Pr. & 0 & -5 & 2 & -2 & -2  \\   
			\hline
			D.C.A.PP. & 1 & 0 & 1 & 3 & 0  \\   
			\hline
			I.C.A.Pu. & 0 & 0 & 4 & -1 & -1  \\   
			\hline
			I.C.A.Pr. & 1 & -3 & 1 & -4 & -4 \\   
			\hline
			I.C.A.PP. & 1 & 1 & 3 & 4 & 1 \\   
			\hline
			
		\end{tabular}
		\caption{Q-factor score for each statement obtained by Zabala's R-package (\cite{Zabala2014}, \cite{Fazio2024}).}\label{conse}
	\end{center}
\end{table}

Through the formulation proposed by Gao and Soranzo, it is possible to reconstruct the characteristics of the various Q-factors. In the their paper, referring to Roger's model (\cite{Rogers1983}), \cite{Fazio2024} identify \textbf{Q-factor 1} as the \textit{Early Majority adopters} category. Conversely, \textbf{Q-factor 5}, perceiving the Metaverse as least significant, aligns with \textit{Laggards}, slowest to adopt due to perceived high costs or risks, preferring community support over individual burden. \textbf{Q-factors 3 and 4} may represent the \textit{Late Majority}, cautious about innovation and adopting after average participants. Q-factor 4 specifically prioritizes content over tools, evaluating the Metaverse on its cultural content delivery. Finally, \textbf{Q-factor 2} aligns with the perspectives of \textit{Innovators and Early Adopters}. 

Q-factor 1 emerges as the primary viewpoint, explaining the most variance in the sample and endorsed by the largest participant group. Participants identified with Q-factor 1 play a pivotal role in shaping tourist destination promotion strategies. Notably, preferences between "Social" and "Advanced" tools lack clarity as distinct entities. Overall, the Metaverse is slightly less favoured than social options. Despite Q-factor 1 participants having the highest Metaverse knowledge, they do not perceive a positive marginal utility compared to social tools, which remain preferred.

Under the assumptions of the Q-methodology and the proposed statement configuration, we can associate the scores of Table \ref{conse} the relative utility values that each Q-factor assigns to each of the 36 proposed strategies.

These findings illuminate the current Metaverse penetration in tourism and stakeholder perceptions of its competitive advantages in destination attractiveness. Despite its potential, imperfect knowledge among key tourism players constrains its integration. While some embrace its novelty, many stakeholders lack a comprehensive understanding, emphasizing its innovative allure. The competition among the various Q-factors opens up numerous scenarios that trace different trajectories regarding the implementation of the Metaverse in the near future.

The game that will decide which Q-factor will prevail and, consequently, which final strategy will be preferred is difficult to decipher; what we can do is try to imagine the evolution of this game.

\subsection{Replicator dynamics and evolutionary game-theory model}
\label{model}

Game theory is the part of mathematics that models strategic behavior (\cite{binmore2007}). A symmetric game is defined by $m$ players  (\emph{individuals}), that have to choose among $n$ \emph{strategies} $\mathbf{str}_{i}$, $i=1,\dots,n$, and as a result each one receive a certain \emph{pay-off}: \textit{each player acts to maximize his pay-off function.} A set of strategies in which no rational player has incentives to deviate from his strategy is called a \emph{Nash equilibrium} (\cite{gibbons1992}). It is possible to prove that, if players are allowed to choose their strategies according to certain probability distributions, there is always at least one Nash equilibrium in every game defined as above. 

Classical game theory primarily deals with one-time interactions among players who act perfectly rationally, fully aware of all the details of the game. In contrast, \textit{evolutionary game theory} considers games that repeat over time and involve players chosen randomly from a sufficiently large population. Evolutionary game theory was initially formulated by biologists studying conflicts among animals and the theory of evolution. In this context, some evolutionary processes modify the types of behavior and their distribution within the population (\cite{Smith1974}).

While classical game theory assumes that players are perfectly informed and rational, evolutionary game theory recognizes that interactions occur in a dynamic context, with individuals who can adapt and change strategies over time. This approach considers how game strategies spread and stabilize in a population, influenced by evolutionary mechanisms such as natural selection and mutation. Consequently, evolutionary game theory offers a more realistic perspective on how behaviours evolve and adapt in natural environments, as well as how these processes can be applied to social and economic contexts, providing a deeper understanding of the dynamics of interaction among individuals.

In evolutionary game theory, it is customary to replace the rationality assumption with a given dynamics. In particular, a population of individuals of $n$ different types where each type adopts one of the strategies $\mathbf{str}_{i}$, $i=1,\dots,n$ is considered. 

The fraction of individuals of type $i$ at time $t\ge 0$ is given by $x_{i}(t)\in[0,1]$, and the \emph{state} of the population is given by $\mathbf{x}(t)=(x_1(t),\dots,x_n(t))\in  S^{n-1}:=\{\mathbf{z}\in\R^{n}_{+}, \sum_{i=1}^nz_i=1\}$, the $n-1$-dimensional \emph{simplex}, for all $t\ge0$.

Payoff functions, often referred to as \emph{fitness} in literature (\cite{Ferriere2011}), are smooth functions denoted as $\psi_{i}:S^{i-1}\to\R$ for $i=1,\dots,n$. Specifically, at a given state $\mathbf{x}$, individuals of type $i$ have a fitness represented by $\psi_{i}(\mathbf{x})$. 

The state of the population evolves according to a specified dynamic, with the \emph{replicator dynamics} being the most widely recognized(\cite{Taylor1978}).

\textbf{Replicator dynamics} is a fundamental concept in evolutionary game theory, providing a mathematical framework to model how the composition of strategies in a population evolves over time (\cite{Nagylaki2013}). Rooted in Darwinian principles of natural selection, replicator dynamics posits that strategies yielding higher payoffs, or fitness, increase in frequency, while those with lower payoffs diminish (\cite{Smith1974}). In more detail, the replicator equation illustrates that the growth rate of a strategy's frequency is proportional to the difference between the strategy's payoff and the average payoff of the population. If a particular strategy's payoff exceeds the population average, its frequency will grow, promoting the spread of advantageous traits. Conversely, if the payoff is below average, the strategy's prevalence will decline. This process leads to the natural selection of optimal strategies over time (\cite{Hansen2024}).

This dynamic is described by differential equations that capture the rate of change in the proportion of each strategy within the population: 

\begin{equation}
	\label{repli}
	\dot{x_i}=x_i\left(\psi_{i}(\mathbf{x})-\overline{\psi}(\mathbf{x})\right),
\end{equation}

where $\overline{\psi}(\bx):=\sum_{i=1}^nx_i\psi_{i}(\bx)$ is the average pay-off.

In this case we have always considered symmetric conflicts, in the sense that all players are indistinguishable. However, many conflicts are asymmetric and it can result in differing gains depending on the species. Let us restrict ourselves to games between two different species, say \(X\) and \(Y\), and always assume that the possible strategies are finite; moreover, let us assume that the two players facing each other always belong to different species.

Let \(E_1, \ldots, E_n\) be the possible strategies for population \(X\) and \(F_1, \ldots, F_m\) the possible strategies for population \(Y\). We denote by \(x_i\) the frequency of \(E_i\) within \(X\), and by \(y_j\) the frequency of \(F_j\) within \(Y\); thus, the state \(x\) of population \(X\) will be a point in \(S_n\), and the state \(y\) of population \(Y\) will be a point in \(S_m\). If an individual using strategy \(E_i\) plays against an individual using strategy \(F_j\), we call \(a_{ij}\) the payoff for the first player, and \(b_{ji}\) the payoff for the second; thus, the game is described by the payoff matrices \(A = (a_{ij})\) and \(B = (b_{ji})\). 

The expressions \(x^T Ay\) and \(y^T Bx\) represent the average payoffs for populations \(X\) and \(Y\) respectively.

As we have done before, we can associate a system of differential equations with the game: we assume that the growth rate \(\frac{\dot{x_i}}{x_i}\) of the frequency of strategy \(E_i\) is equal to the difference between the payoff \((Ay)_i\) and the average payoff \(x^T Ay\). This, and the corresponding hypothesis for the growth rate of the frequency of strategy \(F_j\), lead to the replication system
\begin{equation}
	\dot{x_i} = x_i((Ay)_i - x^T Ay), \quad i = 1, \ldots, n
\end{equation}
\begin{equation}
	\dot{y_j} = y_j((Bx)_j - y^T Bx), \quad j = 1, \ldots, m
\end{equation}
on the space \(S_n \times S_m\) of all states for \(X\) and \(Y\).

These differential equations can be solved simultaneously using numerical algorithms that allow the reconstruction of the temporal trends of the variables under study; however, to do this, it is necessary to set the initial conditions to start from. These conditions become crucial because the form of the differential equations is such that initial conditions with null values for some variables prevent them from evolving.

\subsubsection{Strategies evolution in Stakeholder perspective}
\label{modelstra}
We can apply the general model to our case by hypothesizing that we can schematize the relationship of individual stakeholders who confront (compete with) the decisions of the community in a general sense. 

Specifically, referring to what was proposed in \cite{Fazio2024}, we can imagine classifying stakeholders based on different Q-factors, with each stakeholder deciding among the various Q-factors by choosing based on the expected utility that a particular strategy imposed by the community will provide. From the community's perspective, it can choose from a set of strategies the one that maximizes the expected utility of the collective stakeholders.

For the population of stakeholders (\(X\)), the possible strategies are \{Q-fact-1, Q-fact-2, Q-fact-3, Q-fact-4, Q-fact-5\}, and the frequencies \(x_{qi}(t)\) represent the fractions of the population that \textit{align} with the various q-factors;  \(x_{qi}(t)\) at \(t=0\) will be determined by the number of stakeholders flagged for each Q-factor relative to the total sample (see Table \ref{loadings}).

\begin{table}[htbp!]
	\begin{center}
		
		\begin{tabular}{|c|c|c|c|c|c|}
			
			\hline
			&  Q-fact-1 & Q-fact-2 & Q-fact-3 & Q-fact-4 & Q-fact-5 \\
			\hline
			\(x_{qi}(0)\) & 0.35 & 0.15 & 0.15 & 0.20 & 0.10 \\
			\hline
		\end{tabular}
		\caption{Initial values for \(x_{qi}(t=0)\); the values have been calculated according to to the fraction of stakeholders assigned to the Q-factors (see Table \ref{loadings}).}\label{xqi0}
	\end{center}
\end{table}

On the other hand, for the community \(Y\), the possible strategies correspond to the 36 combinations of the variables analysed indicated by the statements according to the coding as reported in Table \ref{codifica}. The frequencies \(y_i\) correspond to the fraction of stakeholders who assigned the highest value to the \(i\)-th combination during the q-sorting phase.

Regarding the \(y_j(0)\), given that the number of stakeholders is less than the statements, this means that some \(y_j(0)\) could be zero, blocking some evolution paths from the outset. 

To avoid this, we decided to generate for each of the 36 statements 30,000 random values based on a normal distribution with mean and sigma as shown in the Figure \ref{figura2}. The frequencies \(y_j\) at time 0 correspond to the fraction of the 30,000 sequences that assigned the highest value to the \(j\)-th combination.
The obtained values are shown in Table \ref{yi0} (the simulations were repeated multiple times to ensure the robustness of the results obtained).\\

\begin{table}[htbp!]
	\begin{center}

		\begin{tabular}{|c|c|c|c|c|c|}
			
			\hline
			D.R.T.Pu. & D.R.T.Pr. & D.R.T.PP. & I.R.T.Pu. & I.R.T.Pr. & I.R.T.PP.  \\
			\hline
			0.0058 & 0.003 & 0.0010 & 0.0029 & 0.0002 & 0.0022 \\
			\hline
			D.C.T.Pu. & D.C.T.Pr. & D.C.T.PP. & I.C.T.Pu. & I.C.T.Pr. & I.C.T.PP.  \\
			\hline
			0.0073 & 0.0146 & 0.0007 & 0.0273 & 0.0224 & 0.0108 \\
			\hline
			D.R.S.Pu. & D.R.S.Pr. & D.R.S.PP. & I.R.S.Pu. & I.R.S.Pr. & I.R.S.PP. \\
			\hline
			0.0129 & 0.0054 & 0.0630 & 0.0359 & 0.0317 & 0.0358 \\
			\hline
			D.C.S.Pu. & D.C.S.Pr. & D.C.S.PP. & I.C.S.Pu. & I.C.S.Pr. & I.C.S.PP. \\
			\hline
			0.0314 & 0.0159 & 0.0707 & 0.0359 & 0.0833 & 0.0823 \\
			\hline
			D.R.A.Pu. & D.R.A.Pr. & D.R.A.PP. & I.R.A.Pu. & I.R.A.Pr. & I.R.A.PP. \\
			\hline
			0.0382 & 0.0020 & 0.0442 & 0.0570 & 0.0131 & 0.0373 \\
			\hline
			D.C.A.Pu. & D.C.A.Pr. & D.C.A.PP. & I.C.A.Pu. & I.C.A.Pr. & I.C.A.PP. \\
			\hline
			0.0391 & 0.0222 & 0.0505 & 0.0332 & 0.0285 & 0.0319 \\
			\hline
		\end{tabular}
		\caption{Initial values for \(y_{i}(t=0)\); the values have been calculated according to the fraction of the 30,000 sequences that assigned the highest value to the \(j\)-th combination.}\label{yi0}
	\end{center}
\end{table}

Since the loadings of each stakeholder for the various Q-factors can assume a positive or negative value, we added an additional level by considering that for each Q-factor there are two possible strategies \(-Q_i\) and \(+Q_i\). In this case, for each Q-factor, the payoff matrix corresponds to a \(2 \times 36\) matrix where in the first row the values correspond to the respective z-scores (as shown in Table \ref{conse}) and in the second row to the same values with opposite signs. Considering that there are only two possible choices for each factor, we can refer to the frequencies of each \(+Q_i\), \(z_{qi}\), and write the replication equation as follows:
\begin{equation}
	\label{eqzq}
	\dot{z_{qi}}=z_{qi}(1-z_{qi})\Delta\psi^{Q_i} ,
\end{equation}

where $\Delta\psi^{Q_i} \bydef\psi^{+Q_i}-\psi^{-Q_i} = 2 * \psi^{+Q_i}$.

\(\psi^{+Q_i}\) corresponds to the expected utility value for the \(i\)-th Q-factor from the weighted (by \(y_i\)) mix of strategies of \(Y\).

\begin{equation}
	\label{payoffz}
	\psi^{+Q_i} = \sum_{j} zscore^{Q_i}_j y_j,
\end{equation}
where $zscore^{Q_i}_j$ represent the values of Table \ref{conse}.

If $\psi^{+Q_i}$ is positive, we say that \(+Q_i\) dominates \(-Q_i\) and that \(-Q_i\) dominates \(+Q_i\) otherwise. Note that $z_{qi}(t)\stackrel{t\to\infty}{\longrightarrow} 1$ in the first case and $z_{qi}(t)\stackrel{t\to\infty}{\longrightarrow} 0$ otherwise for all non-trivial initial conditions.

The values of \(z_{qi}(t=0)\), for each \(Q_i\), refer to the fraction of stakeholders' positive loadings (see Table \ref{loadings}); in Table \ref{zqi0} we report the values obtained.\\

\begin{table}[htbp!]
	\begin{center}
		
		\begin{tabular}{|c|c|c|c|c|c|}
			
			\hline
			&  Q-fact-1 & Q-fact-2 & Q-fact-3 & Q-fact-4 & Q-fact-5 \\
			\hline
			\(z_{qi}(0)\) & 0.39 & 0.33 & 0.39 & 0.28 & 0.28 \\
			\hline
		\end{tabular}
		\caption{Initial values for \(z_{qi}(t=0)\); the values have been calculated according to the fraction of stakeholders' positive loadings (see Table \ref{loadings}).}\label{zqi0}
	\end{center}
\end{table}

The payoff matrix \(A\) of \(X\) depends on the z-score matrix and on the frequencies \(z_{qi}\) according to the following formula:
\begin{equation}
	\label{payoffx}
	A \bydef a_{ij}= zscore^{Q_i}_j *(2*z_{qi}-1)
\end{equation}

Thus, for the \(x_{qi}\), the following replication equations are obtained:
\begin{equation}
	\label{replix}
	\dot{x_{qi}} = x_{qi}((Ay)_i - x^T Ay), \quad i = 1, \ldots, 5 \;(Qfactors)
\end{equation}

Regarding \(Y\), the payoff matrix \(B\) corresponds to the transpose of \(A\); so for the \(y_i\), the following replication equations are obtained:
\begin{equation}
	\label{repliy}
	\dot{y_j} = y_j((A^Tx)_i - y^T A^Tx), \quad j = 1, \ldots, 36 
\end{equation}

Thus, we obtain a system of differential equations that includes equations (\ref{eqzq}), (\ref{replix}) and (\ref{repliy}) which can be solved numerically, imposing the initial conditions at \(t=0\) as indicated in Table \ref{xqi0}, \ref{yi0}, \ref{zqi0}.

\section{Results and discussion}
\label{results}

As mentioned earlier, our analysis focused on solving the differential equations defined in (\ref{eqzq}), (\ref{replix}) and (\ref{repliy}). Specifically, our system consists of 5 differential equations for $z_{qi}(t)$, 5 for $x_{qi}(t)$, and 36 for $y_{i}(t)$. In total, we have a system of 46 differential equations and 46 initial conditions provided by the tables \ref{xqi0}, \ref{zqi0}, \ref{yi0}. Our goal is to understand the temporal evolutions of variables $z_{qi}(t)$, $x_{qi}(t)$ and $y_{i}(t)$; this will allow us to determine whether the system evolves towards a solution where one factor dominates over the others (and with what sign), and consequently, which strategy will ultimately prevail.

To solve the system of differential equations, we leveraged the capabilities offered by Wolfram's Mathematica software, specifically utilizing the NDSolve function (\cite{ndsolve}). NDSolve is a fundamental tool in Mathematica for the numerical solution of ordinary differential equations (ODEs) and partial differential equations (PDEs), providing a powerful tool for mathematical analysis and simulation of complex dynamic systems. 

NDSolve is highly flexible and can handle a wide range of differential problems, yielding numerical solutions that can be used for further analysis, simulations, or graphical visualizations. In cases such as ours, where coupled ODE systems are solved, NDSolve can return a list of \textit{\textbf{InterpolatingFunction}}(\cite{interpolatingfunction}) objects corresponding to the different dependent variables. 

In the following sections, the obtained results and the temporal trends of the different components analysed are shown.

\subsection{Q-factors dynamics}
\label{ssec:qfac}
The first interesting result concerns the trend of the $x_{qi}(t)$ components, which, together with those of $z_{qi}(t)$, allows us to understand which Q-factor manages to dominate all the others in the long term and, most importantly, whether the sign is positive or negative.

\begin{figure}[htbp!]
	\centering
	\subfigure[Q-factor-1]{
		\includegraphics[width=0.3\textwidth]{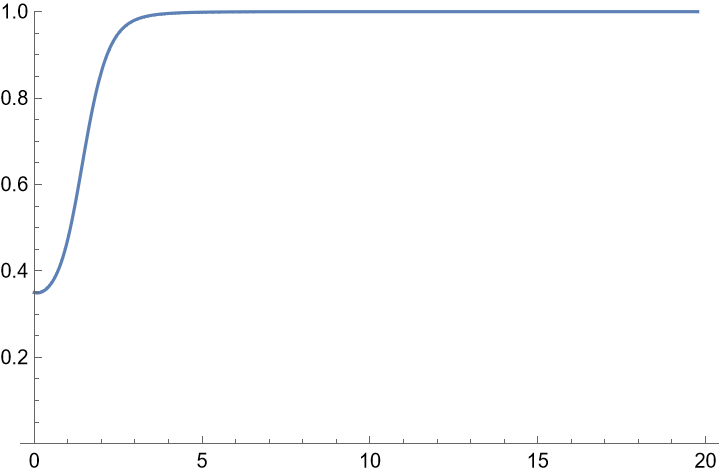}
	}
	\subfigure[Q-factor-2]{
		\includegraphics[width=0.3\textwidth]{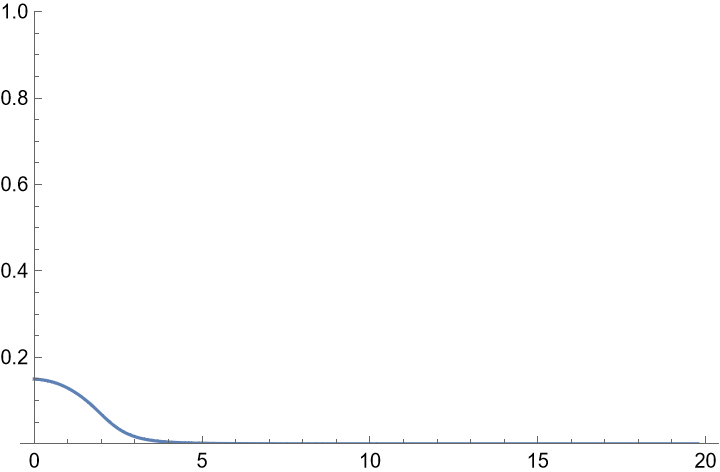}
	}
	\subfigure[Q-factor-3]{
		\includegraphics[width=0.3\textwidth]{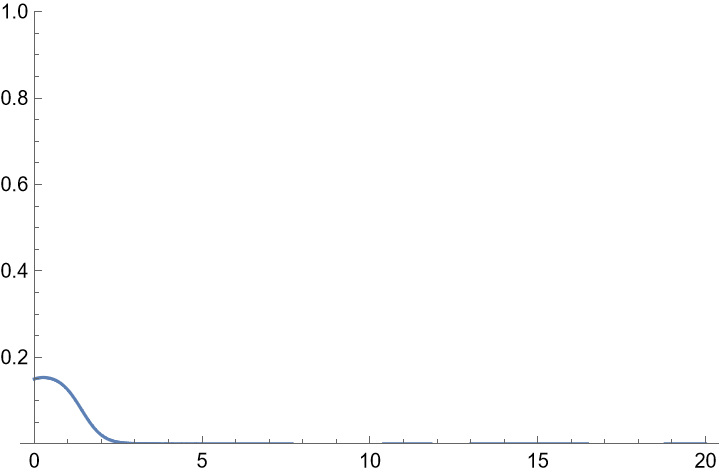}
	}
	\subfigure[Q-factor-4]{
		\includegraphics[width=0.3\textwidth]{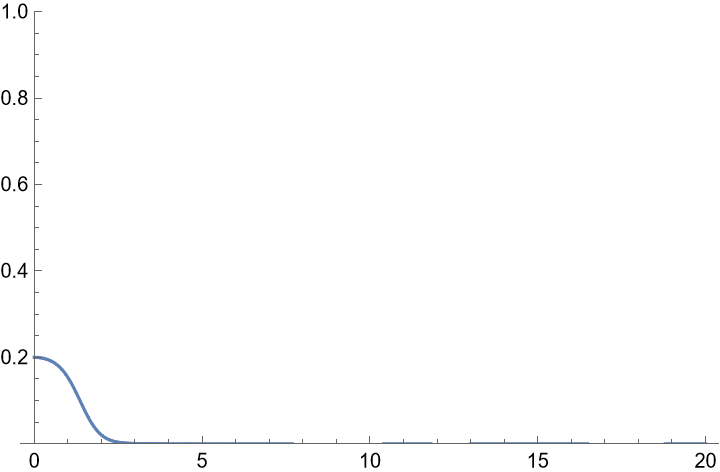}
	}
	\subfigure[Q-factor-5]{
		\includegraphics[width=0.3\textwidth]{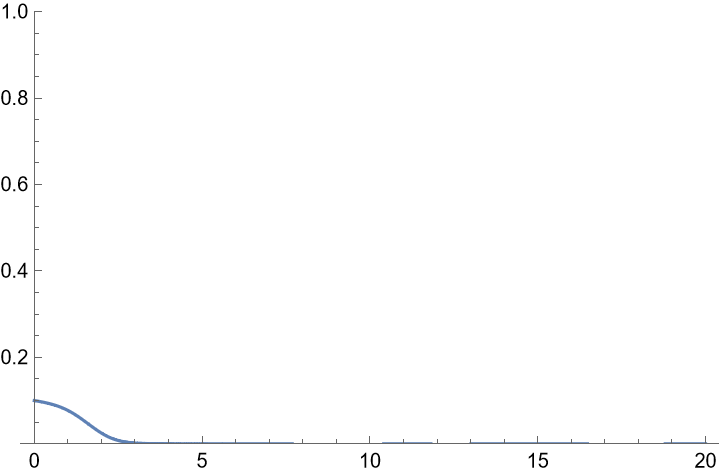}
	}
		\parbox{0.3\textwidth}{
			\vspace{-2,5cm} 
			In the long run, the only surviving factor is Q-factor 1.
		}
	\caption{Trend of the five variables $x_{qi}(t)$ over time, the asymptotic value show that Q-factor-1 dominates over others.}
	\label{xq}
\end{figure}

\begin{figure}[htbp!]
	\centering
	\subfigure[Q-factor-1]{
		\includegraphics[width=0.3\textwidth]{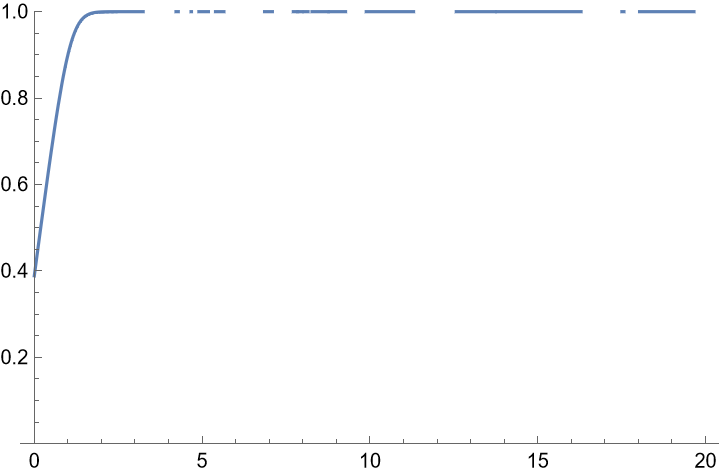}
	}
	\subfigure[Q-factor-2]{
		\includegraphics[width=0.3\textwidth]{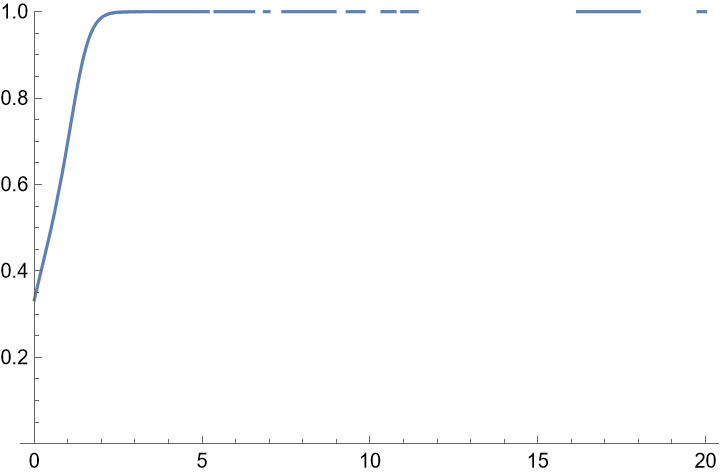}
	}
	\subfigure[Q-factor-3]{
		\includegraphics[width=0.3\textwidth]{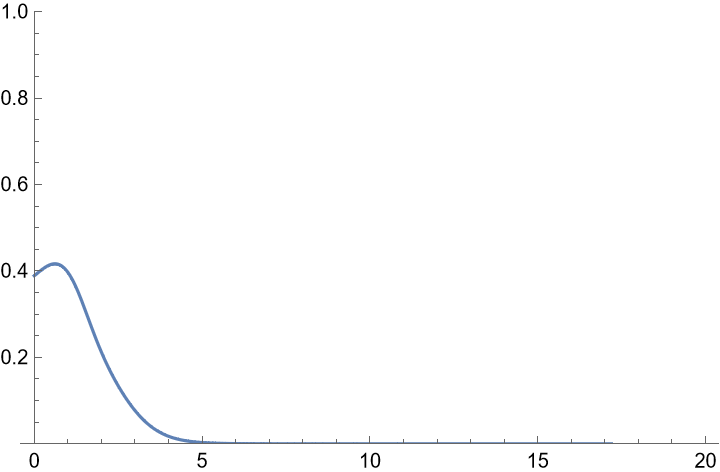}
	}
	\subfigure[Q-factor-4]{
		\includegraphics[width=0.3\textwidth]{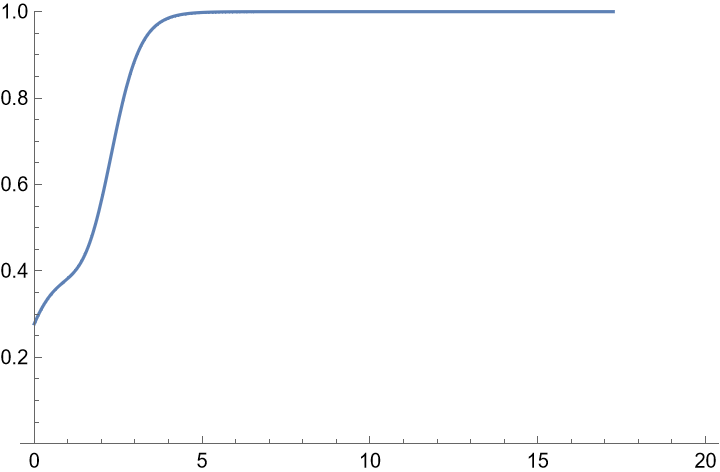}
	}
	\subfigure[Q-factor-5]{
		\includegraphics[width=0.3\textwidth]{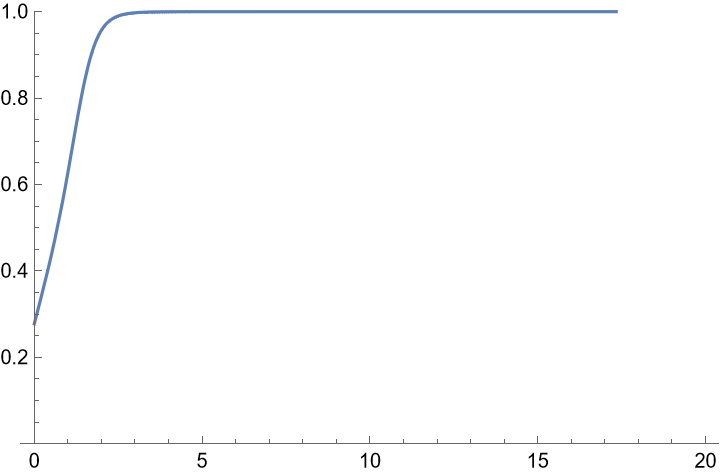}
	}
		\parbox{0.3\textwidth}{
			\vspace{-2,5cm} 
			Please note that $z_{qi}(t)\stackrel{t\to\infty}{\longrightarrow} 1$ in   case \(+Q_i\) dominates \(-Q_i\) and $z_{qi}(t)\stackrel{t\to\infty}{\longrightarrow} 0$ otherwise.
		}
	\caption{Trend of the five variables $z_{qi}(t)$ over time, the asymptotic value determines the main sign of the loading of the various Q-factors. All Q-factors tend to positive values, except for Q-factor 3.}
	\label{zq}
\end{figure}

In figures \ref{xq} and \ref{zq}, the trends of $x_{qi}(t)$ and $z_{qi}(t)$ for all five Q-factors are shown, respectively. The temporal evolution of these variables indicates that the "population" related to \textbf{Q-factor 1} is the one that grows and asserts itself over all the others, with the overall loading tending towards distinctly positive values. From this result, we can deduce that the population will likely stabilize according to the positive Q-factor 1 model, which, as we recall, was identified with the group of innovators defined as \textit{Early Majority}. 

According to Rogers' theory of innovation adoption (\cite{Rogers1983}), "early adopters" are individuals or groups who are among the first to embrace and implement a new technology or change within a social system. They are characterized by their innovativeness, visionary outlook, ability to influence others, and access to resources necessary for adoption. In the context of implementing innovation within an ecosystem, early adopters play a pivotal role. They provide early feedback, enhancing the innovation's development and adaptation to market or community needs. Their credibility and proactive promotion help legitimize the innovation, accelerating its acceptance and adoption by others. Overall, early adopters serve as catalysts for creating demand, fostering a positive cycle of innovation diffusion and integration within broader societal or organizational contexts.

The evidence in \cite{Fazio2024} highlights that currently, \textit{Early adopters} are still in an initial stage of the evolution process, and the critical mass needed to trigger the persuasion process has not yet been reached. Our findings confirm that the diffusion process is in its initial phase but it is likely to evolve towards an increasing presence of such typology of adopters. Their \textit{empirical experience} will be crucial in initiating imitation processes. However, it remains to be understood what the possible strategic evolution could be and which selective/diffusion models might come \textbf{in to the game}.

\subsection{Strategy selection}
\label{stratsel}

\begin{figure}[htbp!]
	\centering
	\subfigure[DRTPu]{
		\includegraphics[width=0.3\textwidth]{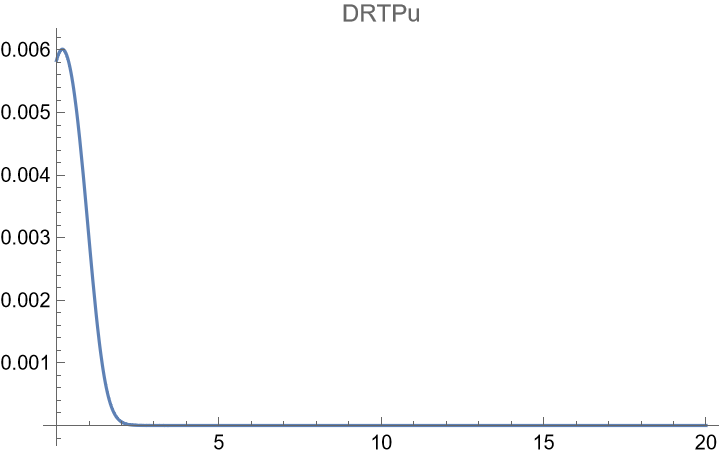}
	}
	\subfigure[DRTPr]{
		\includegraphics[width=0.3\textwidth]{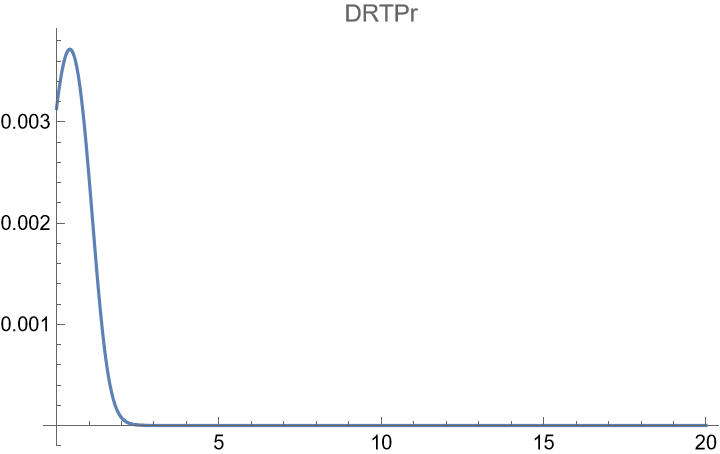}
	}
	\subfigure[DRTPP]{
		\includegraphics[width=0.3\textwidth]{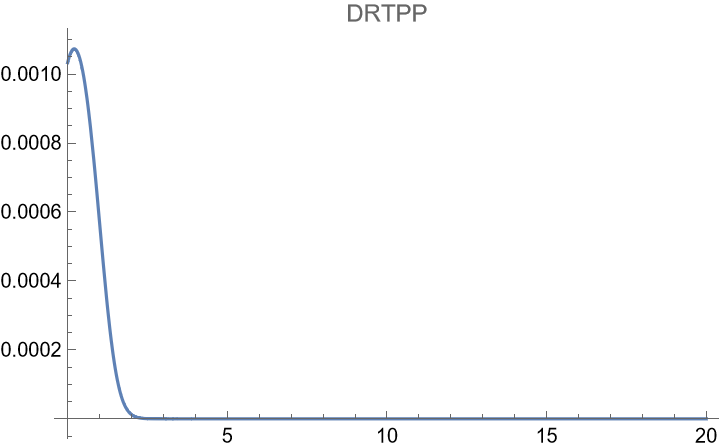}
	}
	\subfigure[IRTPu]{
		\includegraphics[width=0.3\textwidth]{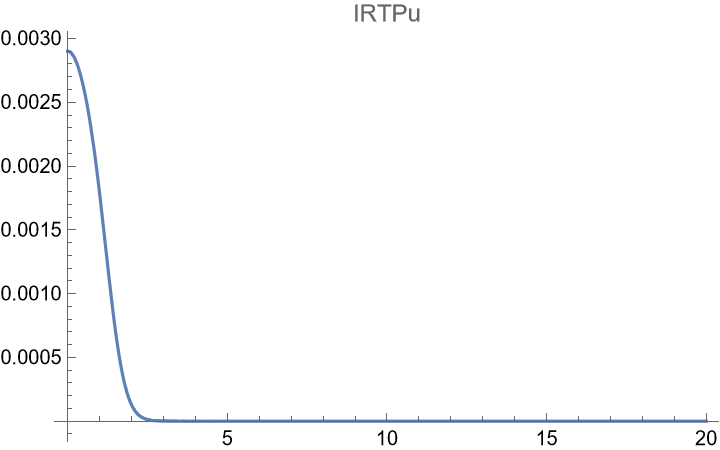}
	}
	\subfigure[IRTPr]{
		\includegraphics[width=0.3\textwidth]{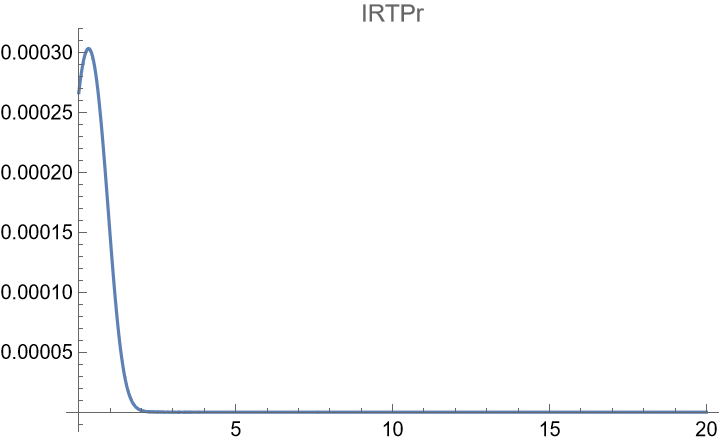}
	}
	\subfigure[IRTPP]{
		\includegraphics[width=0.3\textwidth]{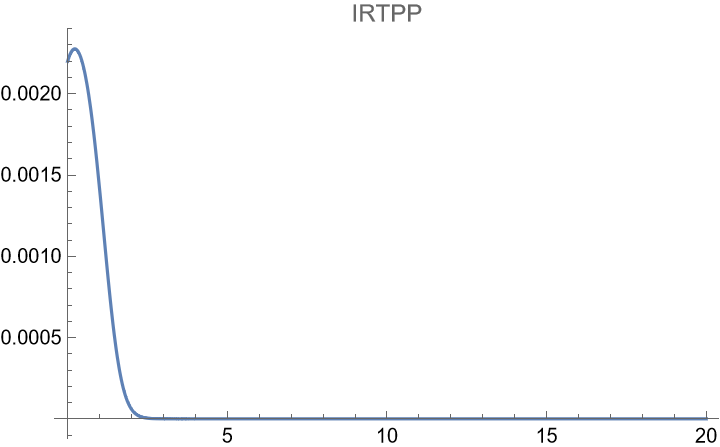}
	}
	\subfigure[DCTPu]{
		\includegraphics[width=0.3\textwidth]{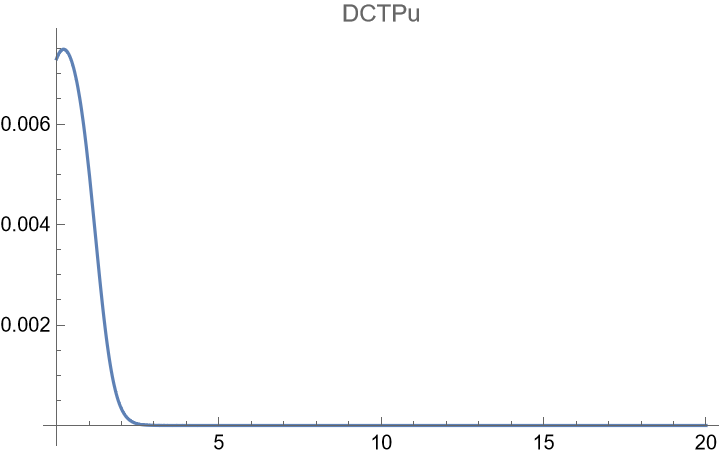}
	}
	\subfigure[DCTPr]{
		\includegraphics[width=0.3\textwidth]{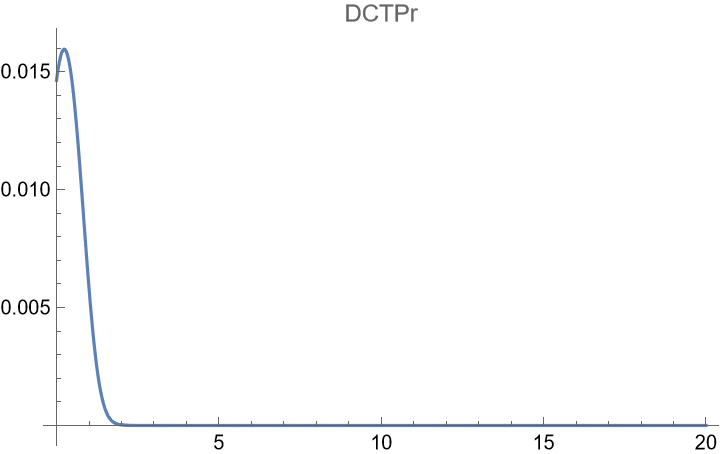}
	}
	\subfigure[DCTPP]{
		\includegraphics[width=0.3\textwidth]{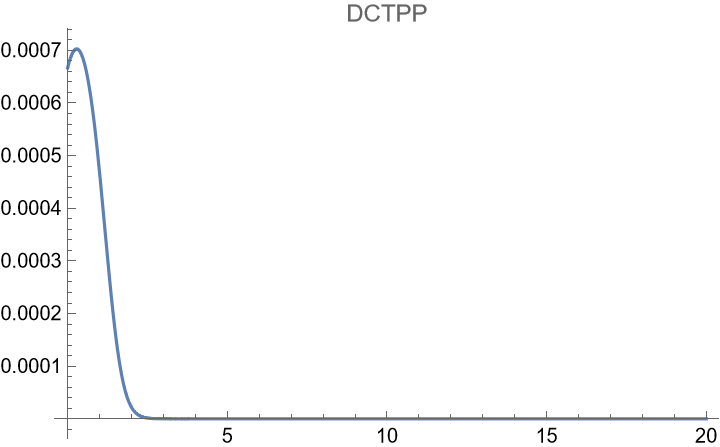}
	}
	\subfigure[ICTPu]{
		\includegraphics[width=0.3\textwidth]{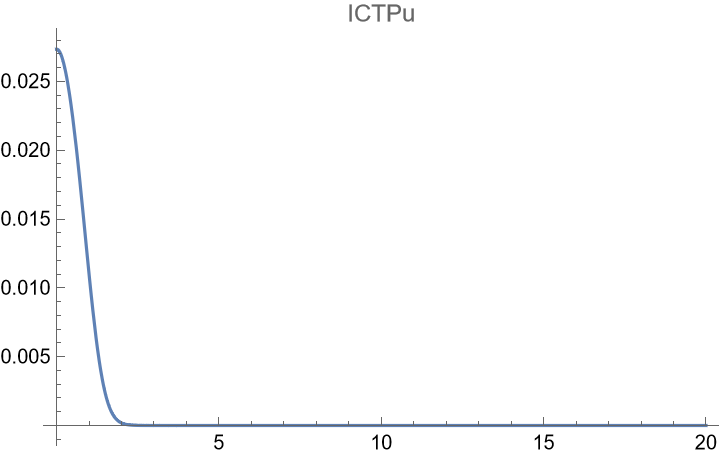}
	}
	\subfigure[ICTPr]{
		\includegraphics[width=0.3\textwidth]{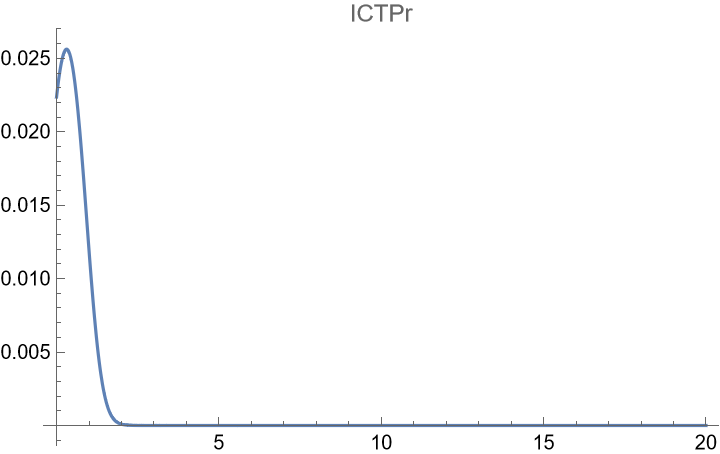}
	}
	\subfigure[ICTPP]{
		\includegraphics[width=0.3\textwidth]{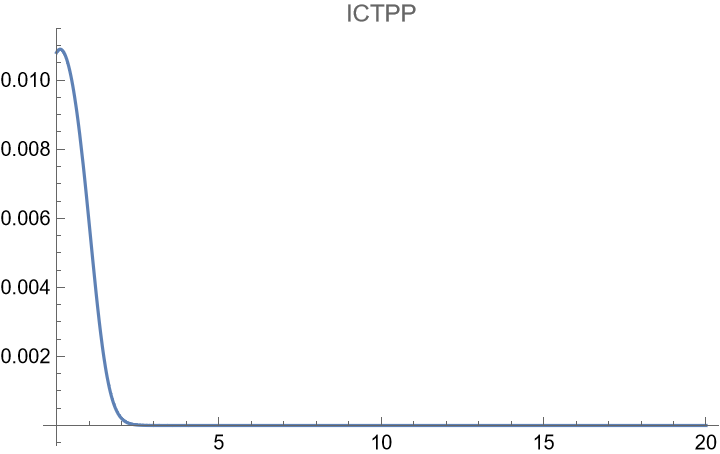}
	}
	\caption{Trends of the variables $y_{i}(t)$ related to the strategies where the Tool dimension takes the \textbf{Traditional (\textit{T})} option.}
	\label{yq1}
\end{figure}

\begin{figure}[htbp!]
	\centering	
	\subfigure[DRSPu]{
		\includegraphics[width=0.3\textwidth]{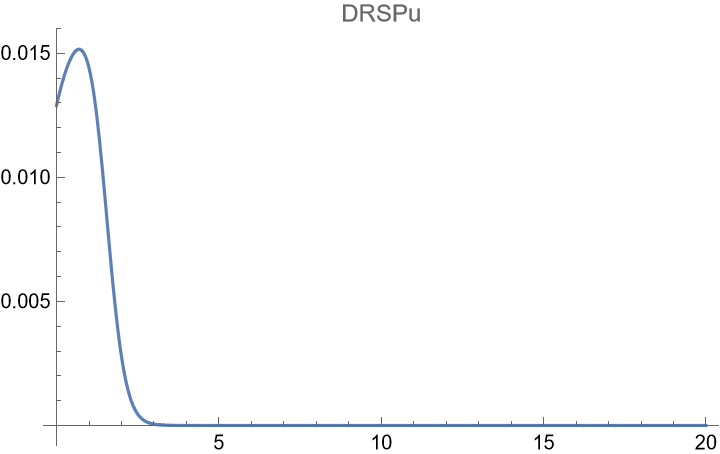}
	}
	\subfigure[DRSPr]{
		\includegraphics[width=0.3\textwidth]{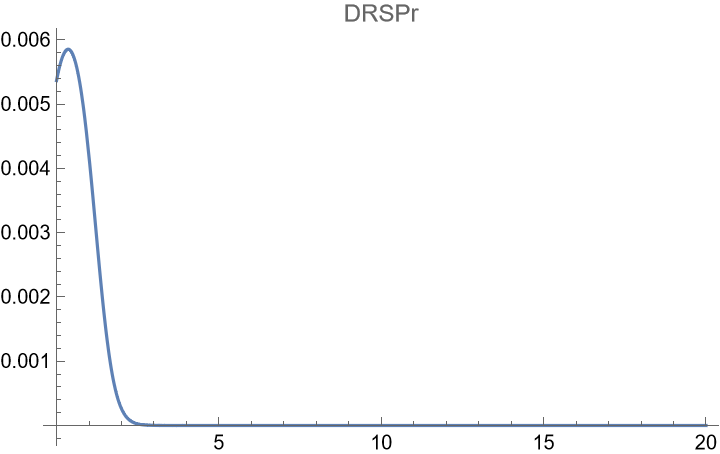}
	}
	\subfigure[DRSPP]{
		\includegraphics[width=0.3\textwidth]{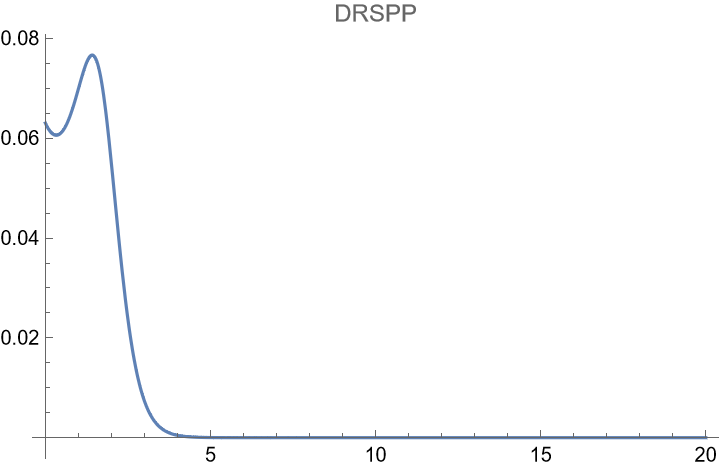}
	}
	\subfigure[IRSPu]{
		\includegraphics[width=0.3\textwidth]{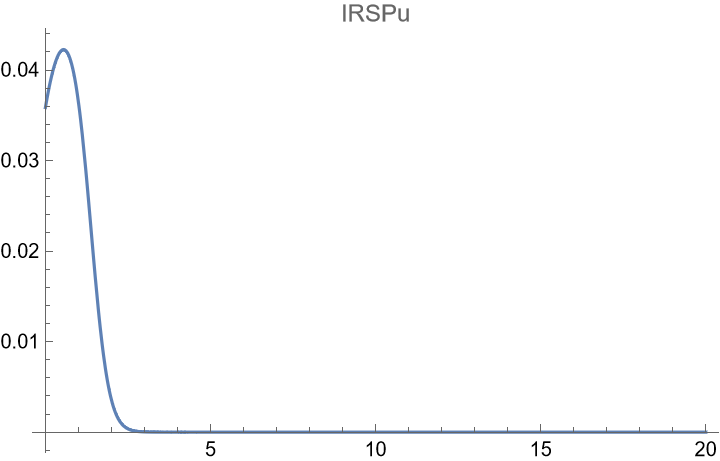}
	}
	\subfigure[IRSPr]{
		\includegraphics[width=0.3\textwidth]{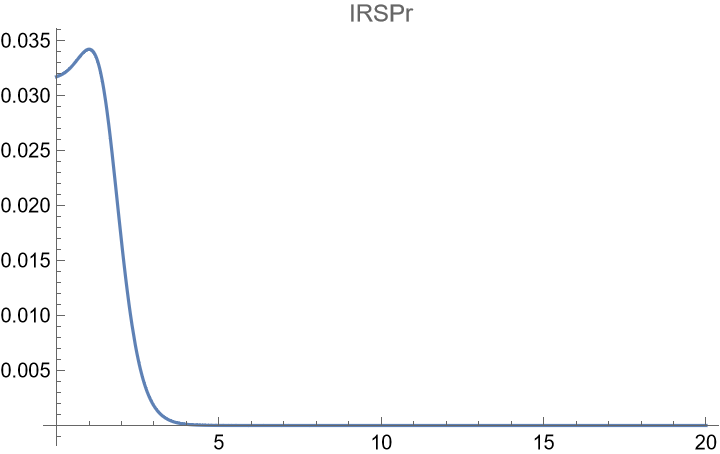}
	}
	\subfigure[IRSPP]{
		\includegraphics[width=0.3\textwidth]{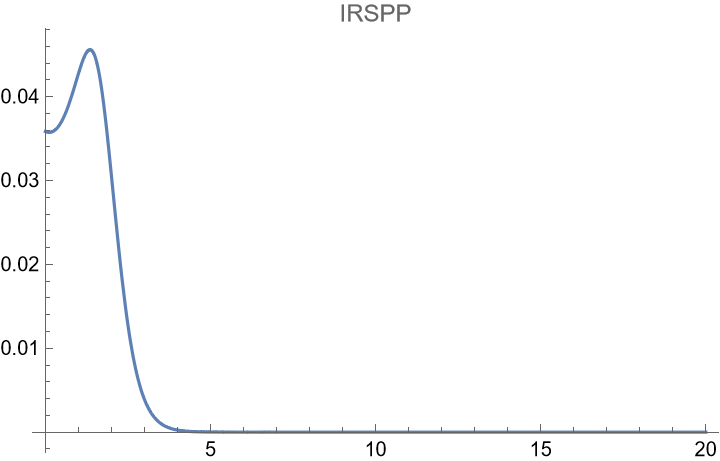}
	}
	\subfigure[DCSPu]{
		\includegraphics[width=0.3\textwidth]{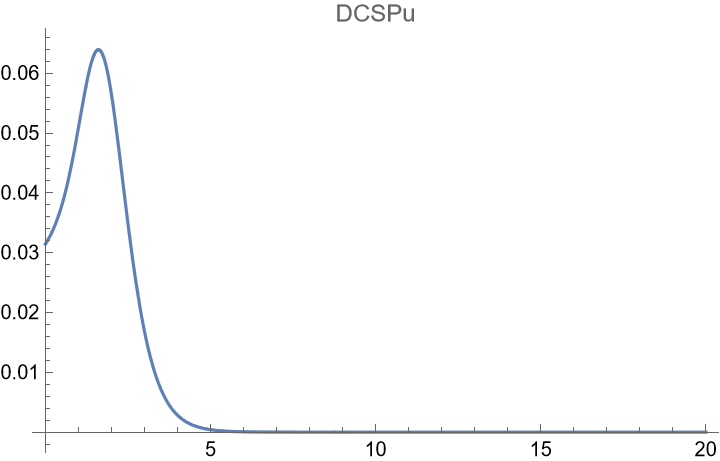}
	}
	\subfigure[DCSPr]{
		\includegraphics[width=0.3\textwidth]{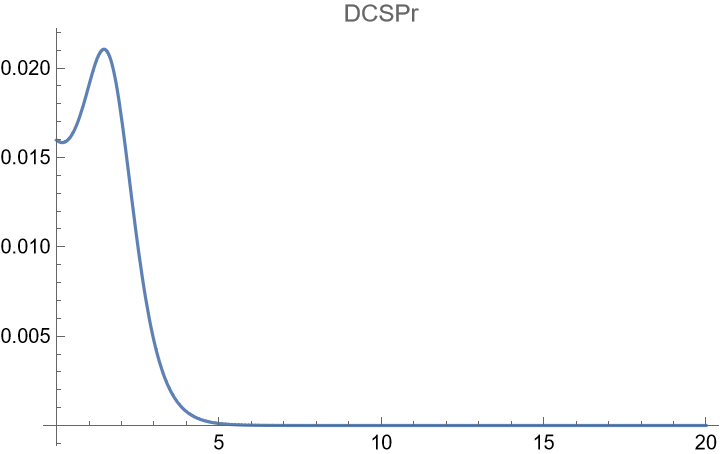}
	}
	\subfigure[DCSPP]{
		\includegraphics[width=0.3\textwidth]{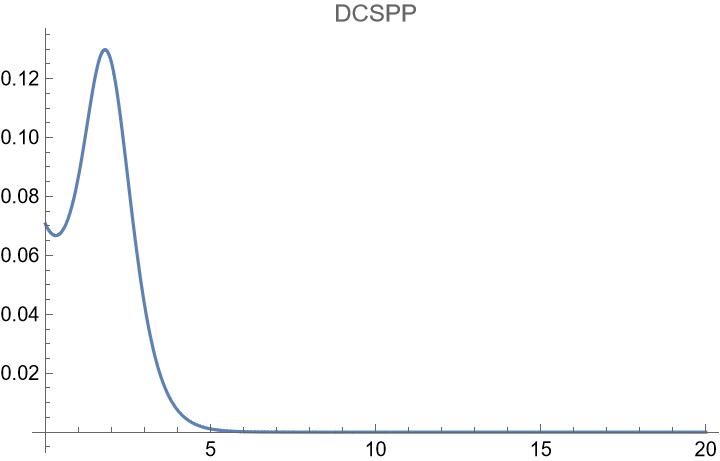}
	}
	\subfigure[ICSPu]{
		\includegraphics[width=0.3\textwidth]{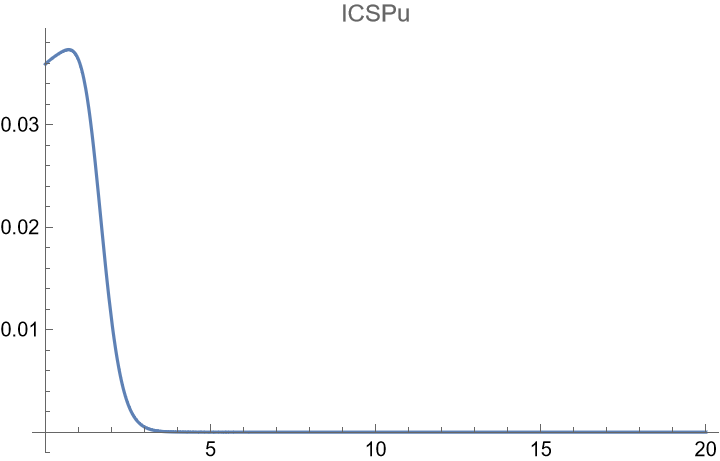}
	}
	\subfigure[ICSPr]{
		\includegraphics[width=0.3\textwidth]{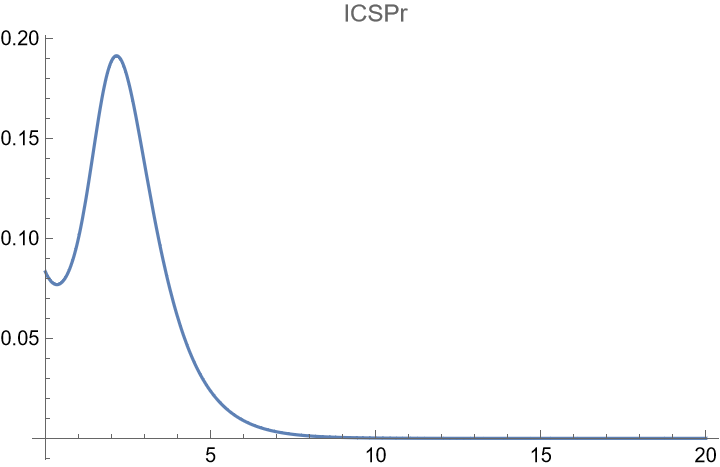}
	}
	\subfigure[ICSPP]{
		\includegraphics[width=0.3\textwidth]{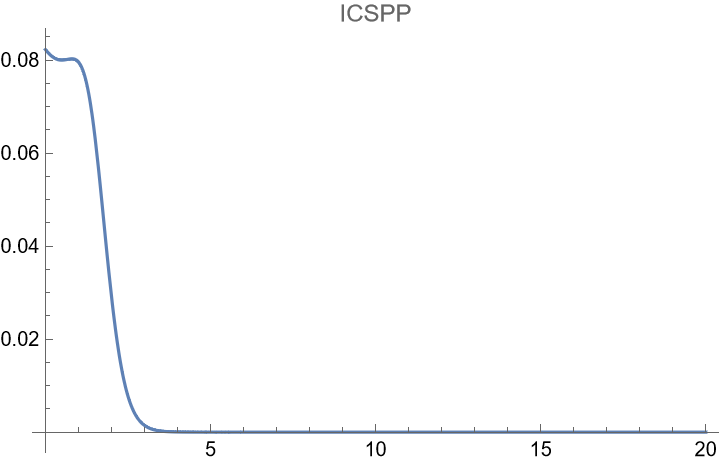}
	}
	\caption{Trends of the variables $y_{i}(t)$ related to the strategies where the Tool dimension takes the \textbf{Standard digital (\textit{S})} option.}
	\label{yq2}
\end{figure}

\begin{figure}[htbp!]
	\centering
	\subfigure[DRAPu]{
		\includegraphics[width=0.3\textwidth]{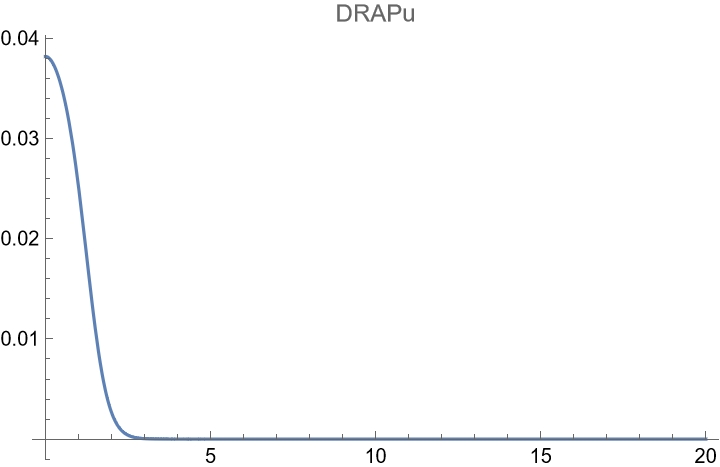}
	}
	\subfigure[DRAPr]{
		\includegraphics[width=0.3\textwidth]{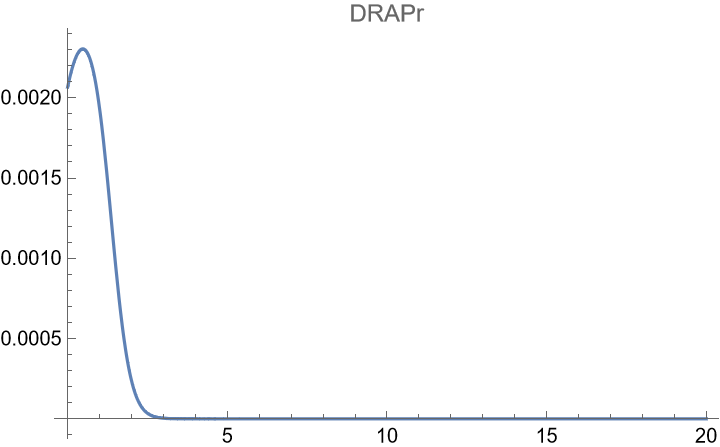}
	}
	\subfigure[DRAPP]{
		\includegraphics[width=0.3\textwidth]{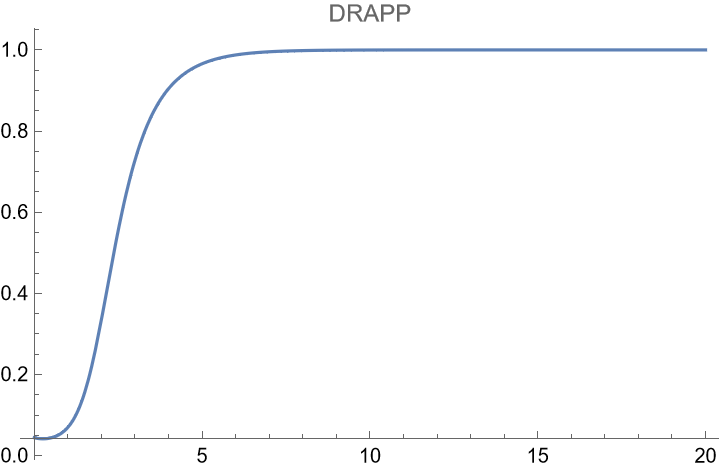}
	}
	\subfigure[IRAPu]{
		\includegraphics[width=0.3\textwidth]{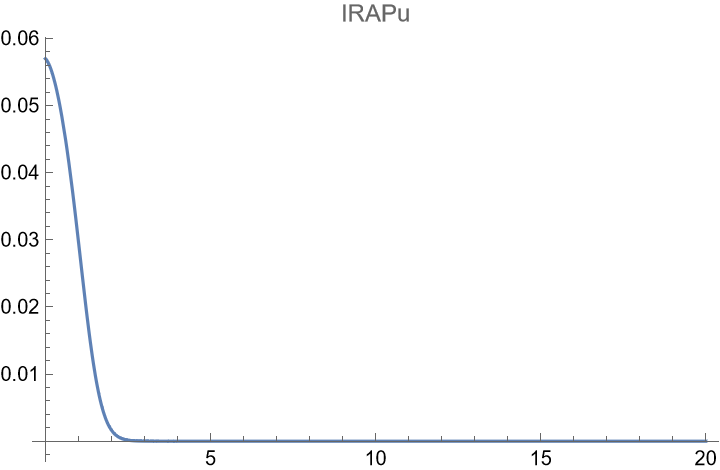}
	}
	\subfigure[IRAPr]{
		\includegraphics[width=0.3\textwidth]{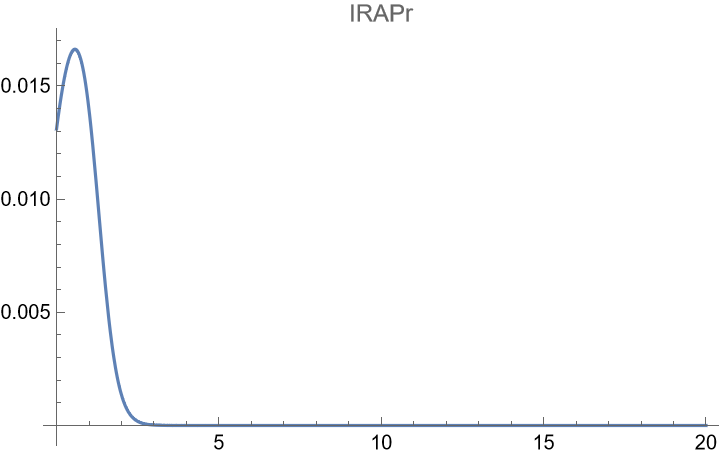}
	}
	\subfigure[IRAPP]{
		\includegraphics[width=0.3\textwidth]{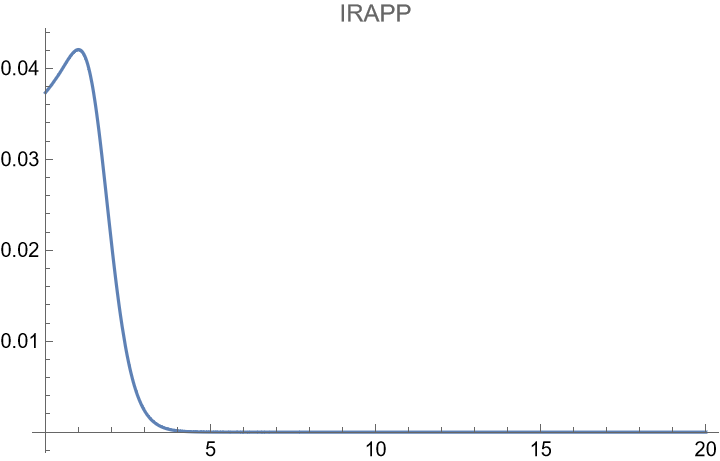}
	}
	\subfigure[DCAPu]{
		\includegraphics[width=0.3\textwidth]{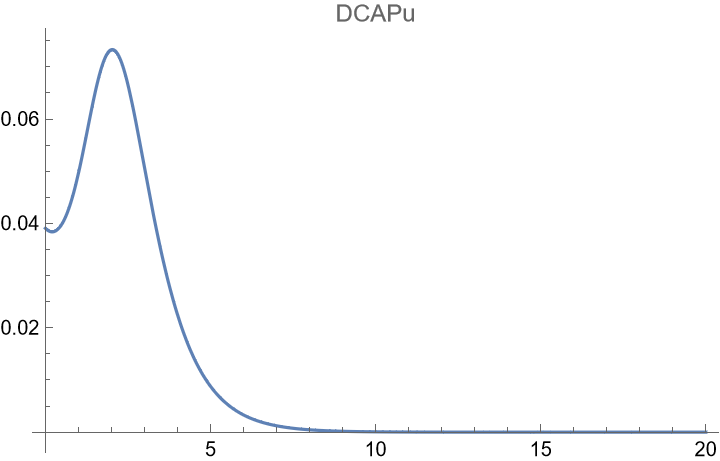}
	}
	\subfigure[DCAPr]{
		\includegraphics[width=0.3\textwidth]{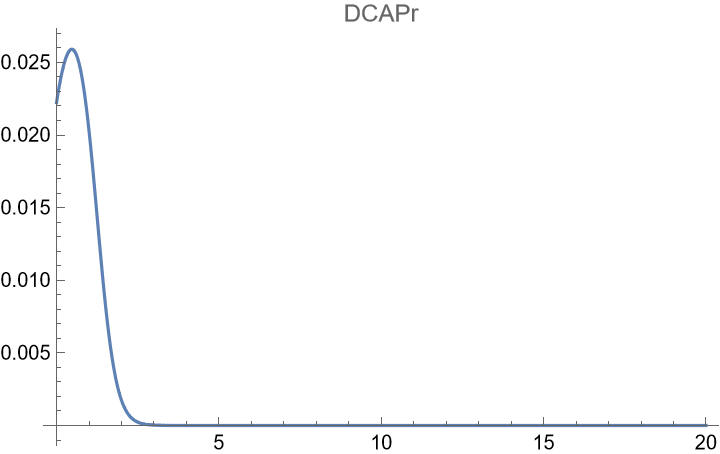}
	}
	\subfigure[DCAPP]{
		\includegraphics[width=0.3\textwidth]{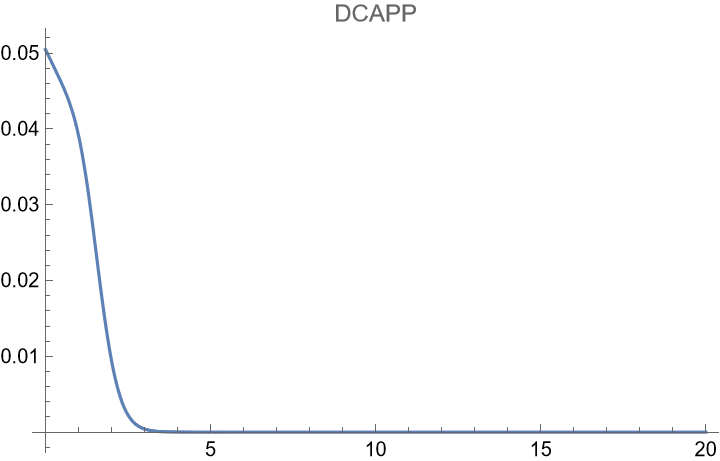}
	}
	\subfigure[ICAPu]{
		\includegraphics[width=0.3\textwidth]{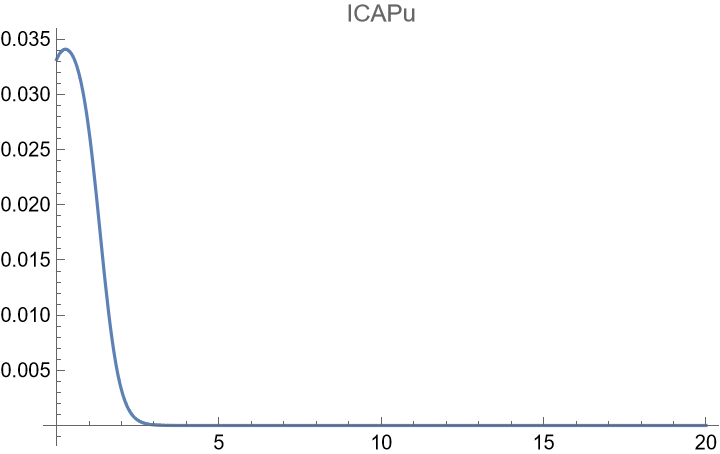}
	}
	\subfigure[ICAPr]{
		\includegraphics[width=0.3\textwidth]{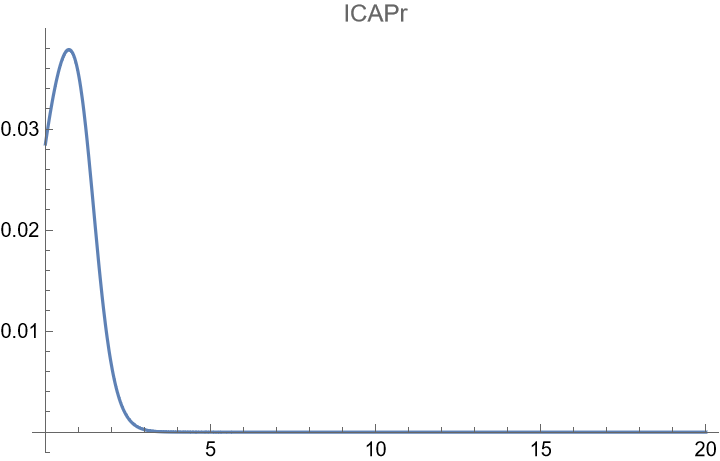}
	}
	\subfigure[ICAPP]{
		\includegraphics[width=0.3\textwidth]{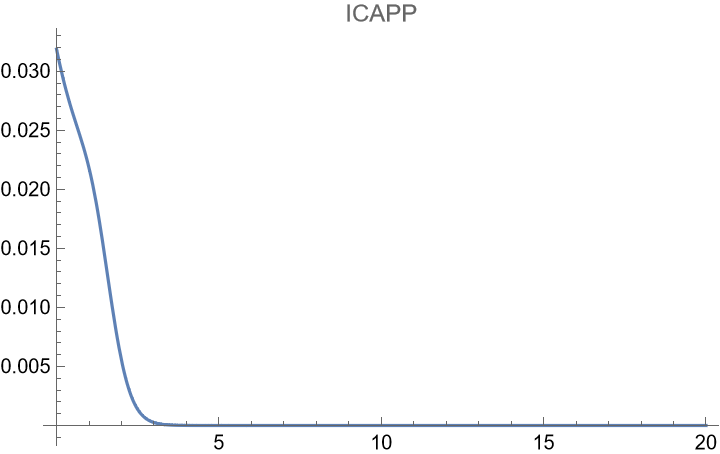}
	}
	\caption{Trends of the variables $y_{i}(t)$ related to the strategies where the Tool dimension takes the \textbf{Advanced digital (\textit{A})} option.}
	\label{yq3}
\end{figure}

The temporal evolution of $x_{qi}(t)$ and $z_{qi}(t)$ is connected to the evolution of the "populations" $y_{i}(t)$ related to the various strategies as defined in Table \ref{codifica}. These populations clearly encompass not only the Tool dimension related to technologies, but also the other three dimensions analysed in the study (see Table \ref{codifica}).  It is important to remember that the conditions imposed at $t=0$ allowed each of the populations of the 36 strategies to be initially non-zero, this gives all of them the possibility to evolve (otherwise, if the initial population for any of the strategies had been zero, it would have remained zero and never "evolved"). The evolution of each of the 36 strategies is shown in Figures \ref{yq1}, \ref{yq2}, and \ref{yq3}. For a better understanding of the evolutionary dynamics, we have divided the 36 strategies into 3 groups based on the \textit{\textbf{Tool}} component. Specifically, in Figure \ref{yq1}, the combinations where the Tool variable takes the Traditional option are grouped; in Figure \ref{yq2}, those with the Social option; and in Figure \ref{yq3}, those with the Advanced option.


Many populations fail to emerge, with values remaining very close to zero. Some populations initially tend to grow but subsequently decrease and tend to nullify over an average time. Only the population of strategy \textbf{DRAPP}, which started from an initial value lower than others (see Table \ref{yi0}), grows continuously and, with a typically sigmoidal trend, imposes itself over all the others.

These latest results depict an evolution of the system that goes through two successive selection phases. In the first phase, some strategies are immediately filtered out while others coexist, generating a mixed strategy that is a weighted combination of the initially surviving strategies (whose populations initially grow). In the subsequent selection phase, one strategy asserts itself over all the others.

\begin{figure}[htbp!]%
	\centering
	\includegraphics[width=0.9\textwidth]{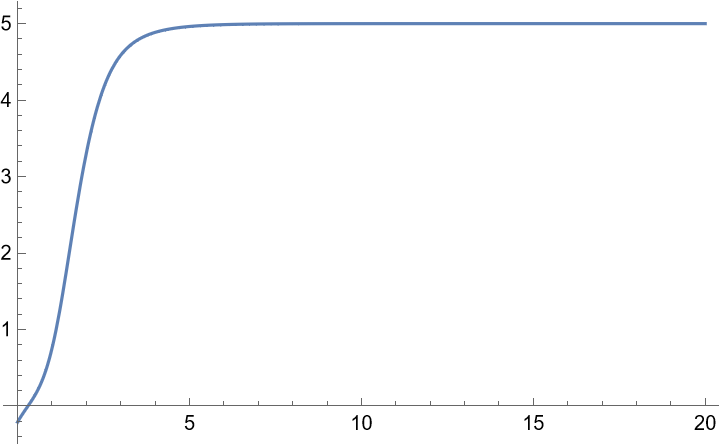}
	\caption{Expected total utility as function of \textit{t}.}\label{totale}
\end{figure}

Lastly, it is also useful to investigate the temporal trend of the expected utility as the evolutionary process follows the different phases of selection. This trend is shown in Figure \ref{totale}. It is interesting to note that the initial expected value, due to the initial mix of different variables, is negative; this value begins to grow initially at a rate that depends on the first phase of selection, where only some strategies are removed and the others are reshuffled into a better mix. Subsequently, the growth rate increases during the second phase of selection, where the system evolves towards a single final solution.

The result of the evolution process thus leads to a selection among the initial mix of strategies, favouring the strategy that involves targeting local "consumers," with recreational content and the use of technologies such as the metaverse, AR (augmented reality), and VR (virtual reality), developed in co-participation between the public and private sectors. This strategy leverages cutting-edge technology to create immersive experiences that can significantly enhance the appeal and accessibility of cultural and tourist attractions.




The potential success of these strategies lies in their ability to provide unique, interactive experiences that resonate with the local population. By focusing on recreational content, these initiatives make cultural heritage more accessible and enjoyable. Moreover, the collaboration between public and private sectors ensures a broad resource base, fostering innovation and sustainability.

\section{Conclusion}

The integration of immersive technologies in cultural and tourism industries is at a critical juncture, driven by stakeholders' perceptions. These technologies promise to revolutionize experiences by offering unprecedented levels of engagement and interaction. For instance, museums and galleries can use AR and VR to create interactive exhibits, enhancing visitors' understanding and enjoyment of art and history. However, high costs, technical complexities, and concerns about accessibility and inclusivity pose significant challenges.

By addressing stakeholders' concerns and leveraging cutting-edge technology, the cultural and tourism industries can create sustainable, impactful strategies that enhance local engagement and preserve cultural heritage. This strategic planning is crucial for staying relevant and competitive in an increasingly digital world.

By employing game theory and Q-methodology, the interactions between various strategies and stakeholders perceptions can be analyze. 

Applying evolutionary game theory models is highly valuable as they provide a dynamic framework for understanding how strategies and behaviours evolve over time within populations. These models are crucial for analysing situations where individuals' success depends not only on their own choices but also on the actions of others. By simulating the adaptation and spread of strategies, evolutionary game theory helps predict long-term outcomes and stability in complex systems, making it an essential tool for studying competition, cooperation, and innovation. Its importance lies in its ability to capture the real-world dynamics of strategic interactions. In this paper, evolutionary game theory has been utilized to anticipate potential evolutionary paths within the economic system centred around cultural heritage and tourism, providing insights into how the ecosystem might evolve in adoption of immersive technologies based on evolutionary principles. 

In the future, it would be beneficial to explore additional case studies to generate alternative evolutionary models and assess differences or similarities. Moreover, incorporating stochastic noise effects or other variations could further refine these predictions and enhance their robustness. 

\section*{Acknowledgement}
.

\bibliographystyle{tfcad}
\bibliography{biblio}

\end{document}